\documentclass[preprint]{revtex4}
\usepackage{bm}

\usepackage[dvips]{graphicx}

\usepackage{color}

\begin{document}

\title{Cooperativity and modularity in protein folding\\
{\small (submitted to the Nobuhiko Sait$\hat{\rm o}$ memorial issue of  {\it Biophysics and Physicobiology})}
}
\author{Masaki Sasai}
\email{sasai@tbp.nuap.nagoya-u.ac.jp}
\affiliation{Department of Computational Science and Engineering and Department of Applied Physics, Nagoya University, Nagoya, Aichi 464-8603, Japan}
\author{George Chikenji}
\affiliation{Department of Computational Science and Engineering and Department of Applied Physics, Nagoya University, Nagoya, Aichi 464-8603, Japan}
\author{Tomoki P. Terada}
\affiliation{Department of Computational Science and Engineering and Department of Applied Physics, Nagoya University, Nagoya, Aichi 464-8603, Japan}

\maketitle
\newpage

\section*{Abstract}
A simple statistical mechanical model proposed by Wako and Sait$\hat{\rm o}$  has explained the aspects of protein folding surprisingly well. This model was systematically applied to multiple proteins by Mu$\tilde{\rm n}$oz and Eaton and   has since been referred to as the Wako-Sait$\hat{\rm o}$-Mu$\tilde{\rm n}$oz-Eaton (WSME) model.  The success of the WSME model in explaining the folding of many proteins has verified the hypothesis that the folding is dominated by native interactions, which makes the  energy landscape globally biased toward native conformation. Using the WSME and other related models, Sait$\hat{\rm o}$ emphasized the importance of the hierarchical pathway in protein folding; folding starts with the creation of  contiguous segments having a native-like configuration and proceeds as growth and coalescence of these segments. The $\phi$-values calculated  for barnase with the WSME model suggested that segments contributing to the folding nucleus are similar to the structural modules defined by the pattern of native atomic contacts. The WSME model was extended to explain folding of multi-domain proteins having  a complex topology, which opened the way to  comprehensively understanding the folding process of multi-domain proteins. The WSME model was also extended to describe allosteric transitions,  indicating that the allosteric structural movement does not occur as a  deterministic sequential change between two conformations but  as a stochastic diffusive motion over the dynamically changing energy landscape. Statistical mechanical viewpoint on folding, as highlighted by the WSME model, has been  renovated in the context of modern methods and ideas, and will continue to provide  insights on equilibrium and dynamical features of proteins.
\\

\noindent 
{\bf Running title:} Cooperativity and modularity \\
\noindent 
{\bf Key words:} WSME model, energy landscape, statistical mechanics\\

\clearpage

\section*{Introduction}
Understanding protein folding is a fascinating problem of biomolecular self-organization, and it is a prerequisite for comprehending the reactions and interactions of proteins. An important method for delineating the folding problem is through  a simple statistical mechanical model. The model was  proposed by  Wako and Sait$\hat{\rm o}$ in 1978 \cite{Wako78a,Wako78b} by extending classical models of helix-coil transitions \cite{Lifson61, Poland66} to many-bodied  heterogeneous cases. However, the model was not widely accepted until quantitative comparison  between the model results and the experimental data became possible. 

Around 1990--2000, three important advances changed the researchers' viewpoint. The first advance was the progress in statistical mechanics of complex systems such as spin glasses and neural networks. Accordingly, a complex system's behavior could be described as a  competition between its tendency  to be trapped into one of extensively many disordered states and its tendency to globally drift  along the energy landscape toward an ordered functional state. Applying this notion to protein folding revealed that the global structure of the folding energy landscape is a key  to explaining the experimental results  \cite{Bryngelson95}. The second advance was the experimental observation of the folding rates of systematically derived mutant proteins, which led to the  $\phi$-value analysis technique to reveal  structures of the transition state ensemble of folding \cite{Fersht99,Fersht00}. The third advance was the drastic increase in computational power, which facilitated not only  large-scale simulations with realistic models but also the quick and accurate evaluation of folding mechanisms with simplified  models. Combining these advances, theoretical models of the energy landscape of folding were introduced to explain and predict the experimentally observed $\phi$-values and other quantities, which led to the innovative cooperation between theories and experiments and promoted a paradigm shift in folding studies \cite{Onuchic04, Dagget03}. The model developed by Wako and Sait$\hat{\rm o}$ was ``re-discovered'' in 1999 by Mu$\tilde{\rm n}$oz and Eaton \cite{Munoz99}, and  this model has since made a  significant contribution to the advancement in folding studies.

A major advantage of this model is that the partition function can be exactly calculated from the model Hamiltonian \cite{Bruscolin02, Pelizzola05}; the exact calculation allows us to obtain a transparent picture on free-energy landscapes, pathways, and rates of folding. The model was at first criticized as quantitatively invalid \cite{Karanicolas03}. However, such invalidity was due to the particular approximation used in the calculation and the problem disappeared when the exact solution of the model was used. 
Since then, the Wako-Sait$\hat{\rm o}$-Mu$\tilde{\rm n}$oz-Eaton (WSME) model has been widely applied  in calculating pathways  \cite{Henry04, Itoh06, Itoh08, Kubelka08, Nelson08, Yu08, Itoh09, Henry13, Sivanandan13, Inanami14} and kinetics  \cite{Henry04, Zamparo06a, Zamparo06b, Yu08, Inanami14} of folding as well as in explaining mechanical unfolding \cite{Imparato07, Imparato08}, amyloidosis \cite{Zamparo10}, and allosteric transitions and functions \cite{Itoh04a,Itoh04b,Itoh10,Itoh11}. In this review, we discuss the physics behind the WSME model and its applications to folding and other intriguing biophysical problems.

\section*{The WSME Model and Cooperativity}
In the WSME model, a protein conformation is described by a set of Ising-like variables,  $\{m_i\}$.  $m_i =1$, when the dihedral angles of the backbone chain at the $i$th residue have similar values to those in the native conformation, and  $m_i =0$ otherwise. The WSME Hamiltonian is defined by a function of $\{m_i\}$ as
\begin{eqnarray}
H_{\rm WSME}(\{m_i\})=\sum_{i=1}^{N-1}\sum_{j=i+1}^{N}\epsilon_{ij}\Delta_{ij}\prod_{k=i}^j m_k,
\end{eqnarray}
where $N$ is the total number of residues in  the protein and $\Delta_{ij}$ represents the pattern of native contacts: $\Delta_{ij}=1$, when  the residues $i$ and $j$ are in contact in the native conformation and $\Delta_{ij}=0$ otherwise. $\epsilon_{ij}<0$ represents the strength of the attractive native interactions, for which we may use  $\epsilon_{ij}\approx -0.3$ to $-1.5$ kcal$\cdot$mol$^{-1}$ depending on the extent of the atomic contacts between the residues $i$ and $j$ in the native conformation \cite{Inanami14}. The partition function is calculated as
\begin{eqnarray}
Z_{\rm WSME}(n)={\rm Tr}_n\exp\left(- H_{\rm WSME}(\{m_i\})/k_{\rm B}T-\sum_{i=1}^N\sigma_im_i\right).
\end{eqnarray}
Here, $0\le n\le 1$ is an order parameter of folding: $n=0$ when the chain is completely disordered, $n=1$ when the structure is identical to that determined via X-ray or NMR analysis. Tr$_n$ is a sum under the constraint  $M=\sum_{i=1}^Nm_i=Nn$ as ${\rm Tr}_n=\sum_{m_1=0,1}\sum_{m_2=0,1}\cdots\sum_{m_N=0,1}\delta_{M,Nn}$, where $\delta_{M,Nn}$ is a Kronecker delta.
$-\sigma_i$ represents the reduction of entropy upon structure ordering at the residue $i$, and we may use  $\sigma_i k_{\rm B}\approx 2$--3 cal$\cdot$mol$^{-1}$K$^{-1}$ \cite{Inanami14}. From Eq.2, we can calculate the free energy, $F(n)=-k_{\rm B}T\log Z_{\rm WSME}(n)$, which is the one-dimensional free-energy landscape represented as a function of $n$. The expression of Eq.~2 can be easily extended to the two-dimensional version,  $Z_{\rm WSME}(n_1,n_2)$, with the corresponding free-energy landscape, $F(n_1,n_2)$, by introducing the two-dimensional folding order parameter  $(n_1,n_2)$ with $n_1=\sum_{i=1}^{N_1}m_i/N_1$, $n_2=\sum_{i=N_1+1}^{N}m_i/N_2$, and $N_1+N_2=N$ \cite{Itoh06,Itoh08,Itoh09,Inanami14}; the higher-dimensional representation is also feasible \cite{Itoh09}.

The WSME model is based on two major assumptions. First, it does not consider non-native interactions. Since only  native interactions are explicitly considered in Eq.~1,  the energy monotonously decreases as the chain approaches the native conformation, {\it i.e.,} the energy landscape has a global bias toward native conformation. This global bias has been considered as a characteristic of sequences selected by evolution to meet consistency between local and global structures \cite{Go83} or to show minimally frustrated interactions \cite{Bryngelson95}. The model with such a global bias was first considered by G$\bar{\rm o}$ and his colleagues \cite{Taketomi75,Go81,Abe81}, and the WSME model belongs to a class of such ``G$\bar{\rm o}$-like models''.

\begin{figure}[htbp]
\includegraphics[width=8cm]{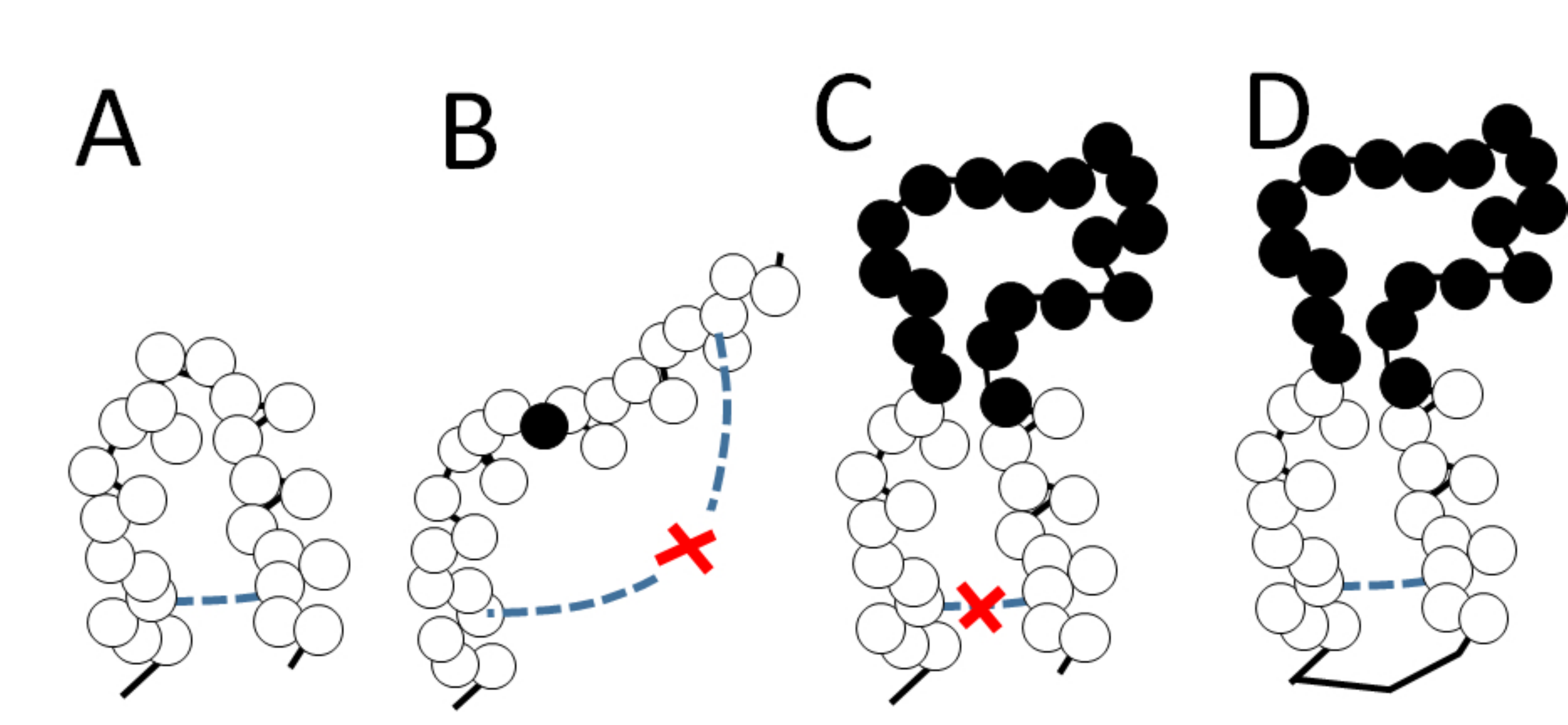}
\caption{
The native interaction  in the WSME model. Residues in the native-like configuration are shown with white circles, and residues in  non-native configurations are shown with filled circles. A) The native interaction (a blue dashed line) between the residues within a contiguous native-like segment is taken into account in the WSME model.  B) The interaction becomes ineffective when an intervening residue is in the non-native configuration. C) If the linker chain connecting two native-like segments is long enough, a number of residues with  random configurations can compensate each other to allow two segments to reach the positions where native interactions are effective. This type of interaction, however, is not taken into account in the WSME model. D) Interactions as in C can be suitably calculated with the WSME Hamiltonian if we consider that the N- and C-termini are connected by a virtual link, as explained in the section ``The WSME Model for Multi-domain Proteins''.
}
\end{figure}

Another significant assumption in the model is that a native interaction occurs only within the ``island'' of a native-like configuration; the $\epsilon_{ij}$ term in Eq.~1 has a nonzero contribution to $H_{\rm WSME}$ only when the consecutive segment from  residues $i$ through $j$ assume  native-like configurations, satisfying $m_i m_{i+1}\cdots m_{j-1}m_j=1$.  This assumption is illustrated in Fig.~1, where intra-segment native interactions are effective (Fig.~1A), but interactions are ineffective when an intervening residue takes the ``wrong'' direction (Fig.~1B). 
This assumption seems plausible when we consider that the residues should  form a local ordered structure through compact atomic packing of residue side chains. Such local structural ordering should be represented  as a  cooperative many-residue correlation given by $m_i m_{i+1}\cdots m_{j-1}m_j=1$ and not as a naive summation of pairwise correlations.

With these two assumptions, contiguous native-like segments are energetically stabilized. Therefore, as illustrated in Fig. 2, folding starts with the creation of short segments with the native-like configuration and proceeds through growth and coalescence of these segments into a larger region to assume the native conformation. We should note that there are combinatorially many ways of segment creation, growth, and coalescence, and  the statistical weight of these different pathways is evaluated with the WSME model to explain the distribution of folding pathways  observed in the ensemble of protein molecules.

\begin{figure}[htbp]
\includegraphics[width=8cm]{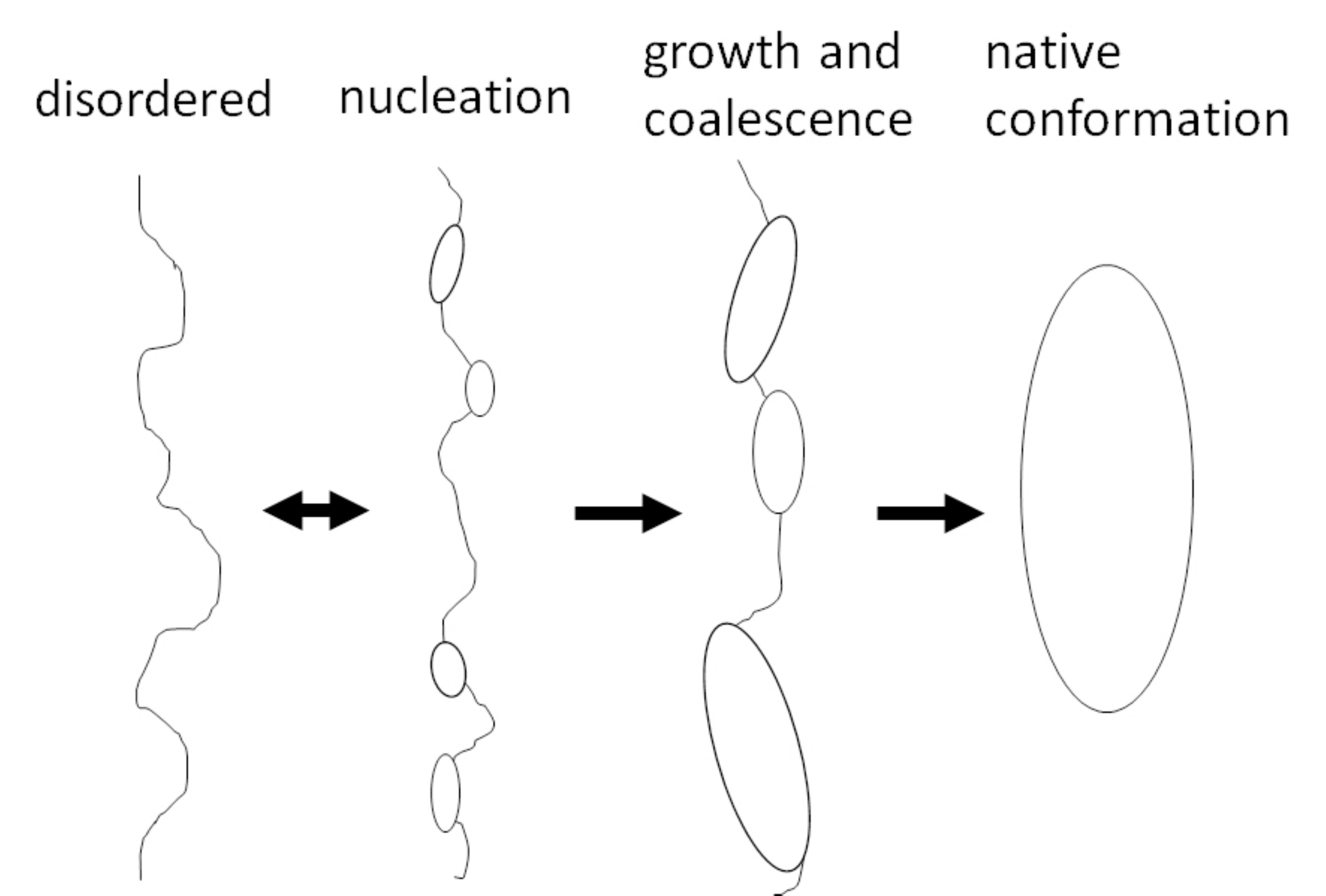}
\caption{
The hierarchical process of  protein folding. Folding starts with the creation of contiguous segments with a native-like configuration. After nucleation, folding proceeds as those segments grow and coalesce into larger regions to reach native conformation.
}
\end{figure}

\begin{figure}[htbp]
\includegraphics[width=10cm]{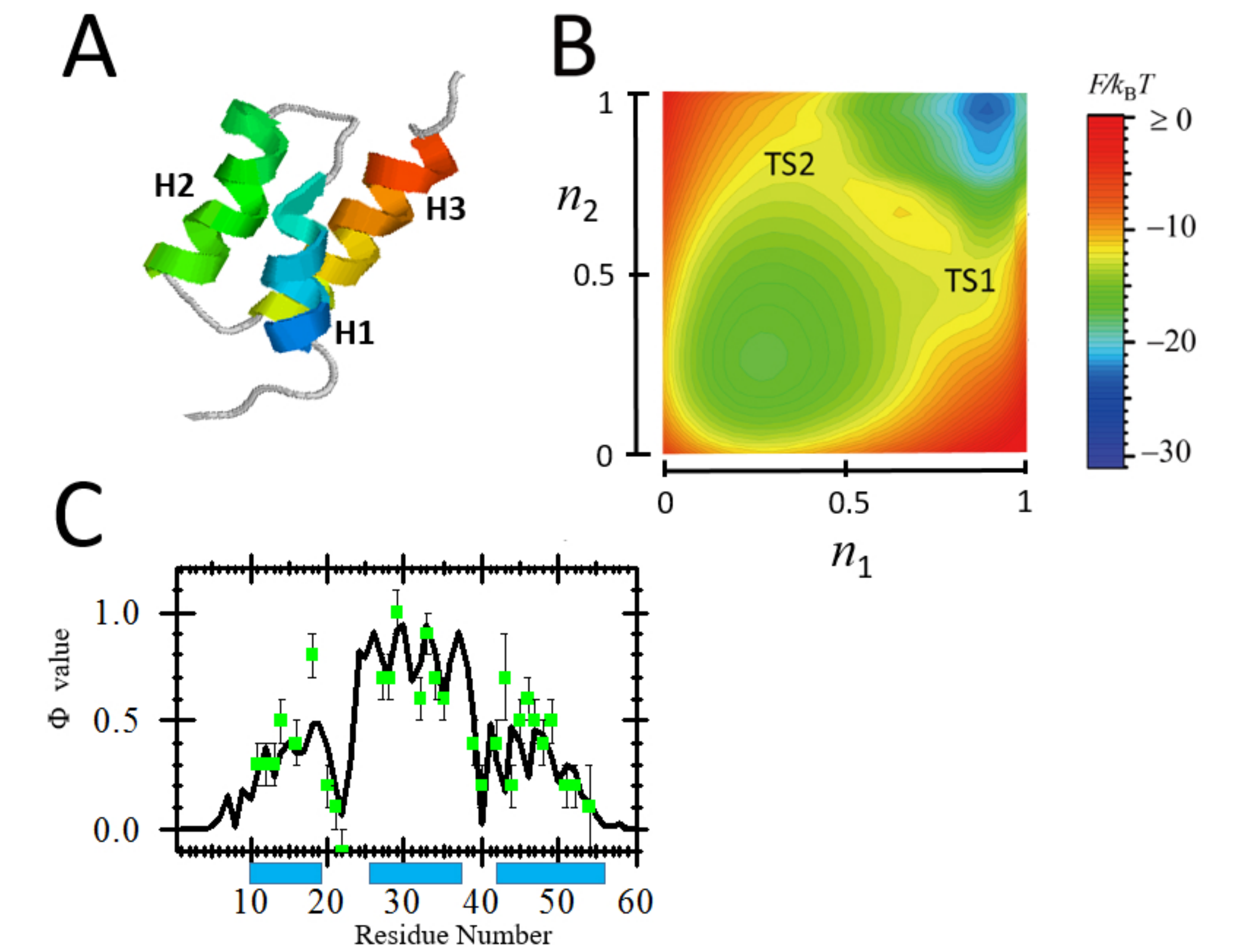}
\caption{
Application of the WSME model to the B domain of {\it Staphylococcal} protein A (BdpA). A) Native conformation of BdpA (Protein Data Bank (PDB) code:~1bdd). B) Two-dimensional free-energy landscape, $F(n_1,n_2)$, calculated with the WSME model, where $n_1$ is the folding order parameter of the N-terminal half, and $n_2$ is the one of the C-terminal half. A contour is drawn every 0.5$k_{\rm B}T$. $F(n_1,n_2)$ has two basins: the unfolded state basin ($n_1\approx 0.3, n_2\approx 0.3$) and the basin of the native state ($n_1\approx 1.0, n_2\approx 1.0$). Two transition states, TS1 and TS2, are shown;  there are two dominant  pathways of folding, which proceed through TS1 and TS2. C) Comparison of the calculated and observed $\phi$-values. The calculated values are shown with a line and the observed values \cite{Sato04} are green squares shown with error bars. Bars on the bottom represent the positions of  $\alpha$ helices. Modified from Figs. 1, 3, and 5 of \cite{Itoh06}.
}
\label{Fig:BdpA}
\end{figure}

The WSME model quantitatively explains free-energy landscapes, pathways, $\phi$-values, and kinetic rates of the folding of various proteins \cite{Henry04, Itoh06, Itoh08, Kubelka08, Nelson08, Yu08, Itoh09, Henry13, Sivanandan13, Inanami14}.  In Fig.~\ref{Fig:BdpA}, an example result is shown for the B domain of protein A (BdpA). As shown in Fig.~\ref{Fig:BdpA}A, BdpA is a small 60 residue, $\alpha$-helical protein comprising three helices: H1, H2, and H3. BdpA demonstrates a two-state folding transition between the unfolded and native states \cite{Sato04}. 
The two-dimensional  free-energy landscape $F(n_1,n_2)$ was calculated, where $n_1$ is the order parameter of  folding for the N-terminal half, and $n_2$ is the one for the C-terminal half.
In $F(n_1,n_2)$ of Fig.~\ref{Fig:BdpA}B,  we find two basins: one at a small  $n_1$ and a small  $n_2$, which corresponds to the unfolded state, and the other at $(n_1, n_2)\approx (0.95,0.96)$, which corresponds to the native state. In this landscape, we find two saddles with similar free-energy heighs; therefore, BdpA has  two dominant transition states, TS1 and TS2, in this representation. Along the pathway through TS1, the helix H1 folds earlier than H3, whereas along the pathway through TS2, H3 folds earlier than H1. 
The $\phi$-values were calculated at TS1 and TS2 with the WSME model. Here, the $\phi$-value represents the frequency of structure formation at each residue in the transition state ensemble. 
By averaging the $\phi$-values at two TSs with the respective weights of the Boltzmann factor, the average $\phi$-values are  calculated and compared with the observed ones in Fig.~\ref{Fig:BdpA}C, which shows good agreement between the calculated and  observed data. The existence of two TSs having almost equivalent free-energy heights is due to the symmetrical native conformation of BdpA, as shown in Fig.~\ref{Fig:BdpA}A, and a subtle difference in the experimental conditions or  settings of the simulation model should  break this symmetry and change the relative heights of TS1 and TS2. The results of several simulation studies are conflicting on which helix, H1 or H3, folds earlier \cite{Wolynes04}, but the WSME model provides  a clear explanation of the reason for this disagreement; a symmetrical native conformation brings about the competing multiple pathways of folding and the detailed  simulation condition or the parameter setting modulates the relative statistical importance of multiple pathways.

\begin{figure}[htbp]
\includegraphics[width=10cm]{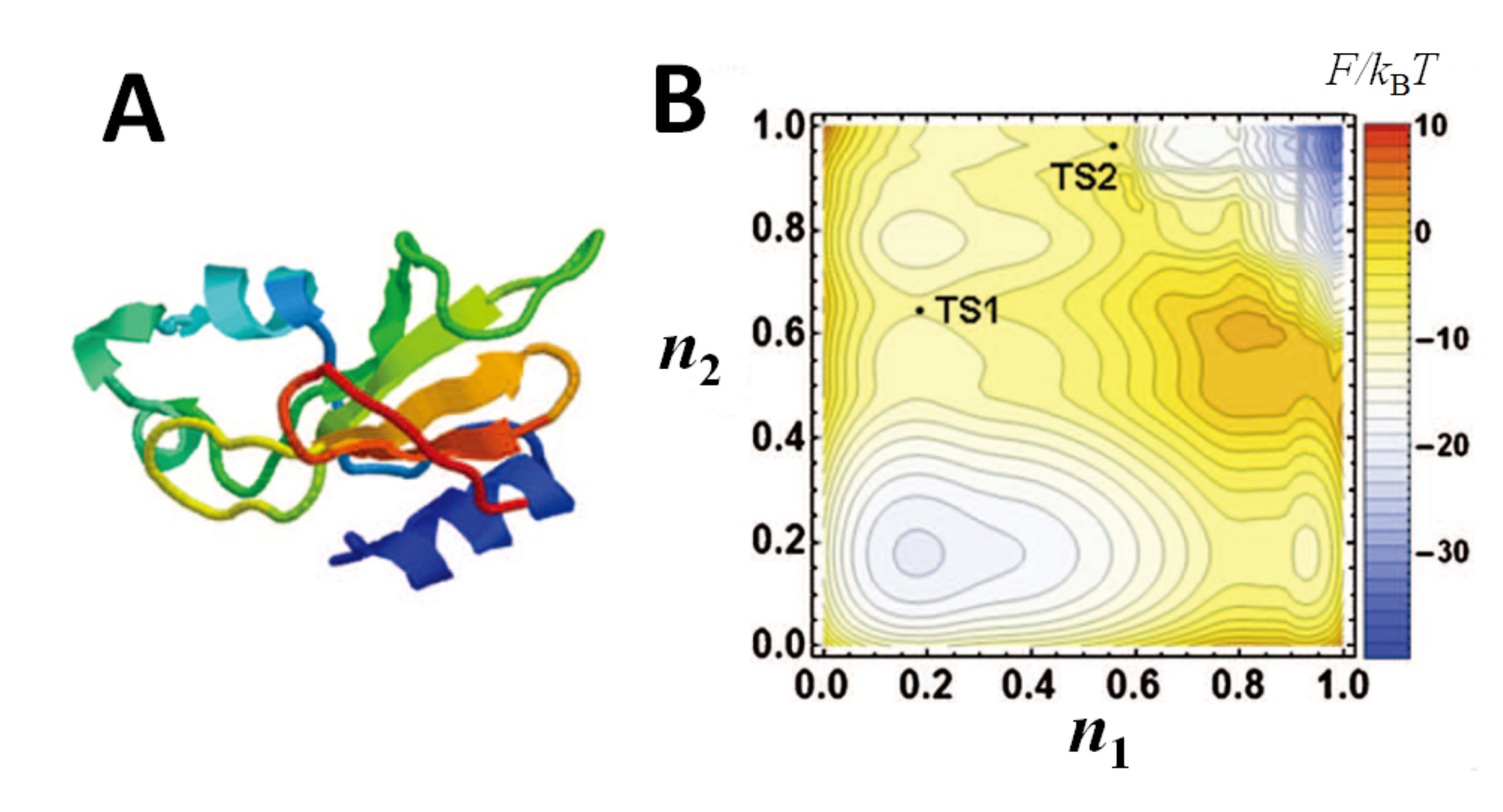}
\caption{
Application of the WSME model to barnase. A) Native conformation of barnase from {\it Bacillus amyloliquefaciens} (PDB code:~1a2p). B) Two-dimensional free-energy landscape, $F(n_1,n_2)$, calculated with the WSME model, where $n_1$ is the order parameter of folding of the N-terminal half, and $n_2$ is the one of the C-terminal half. Contour is drawn in every 2$k_{\rm B}T$. $F(n_1,n_2)$ has four basins; basin of unfolded state ($n_1\approx 0.2, n_2\approx 0.2$), basin of native state ($n_1\approx 1.0, n_2\approx 1.0$), and two basins of intermediate states, I$_{1}$ ($n_1\approx 0.2, n_2\approx 0.8$) and I$_{2}$ ($n_1\approx 0.9, n_2\approx 0.2$). Saddles around the basin I$_{1}$ are much lower in free energy than those around I$_{2}$ are; therefore, a pathway through I$_{1}$ is a dominant pathway, and I$_{1}$ is a dominant intermediate. I$_{2}$ could be detected as an off-pathway intermediate. Along the dominant pathway, there are two transition states, TS1 and TS2. Modified from Fig.~14 of \cite{Itoh09} with permission.
}
\label{Fig:barnase}
\end{figure}

\begin{figure}[htbp]
\includegraphics[width=12cm]{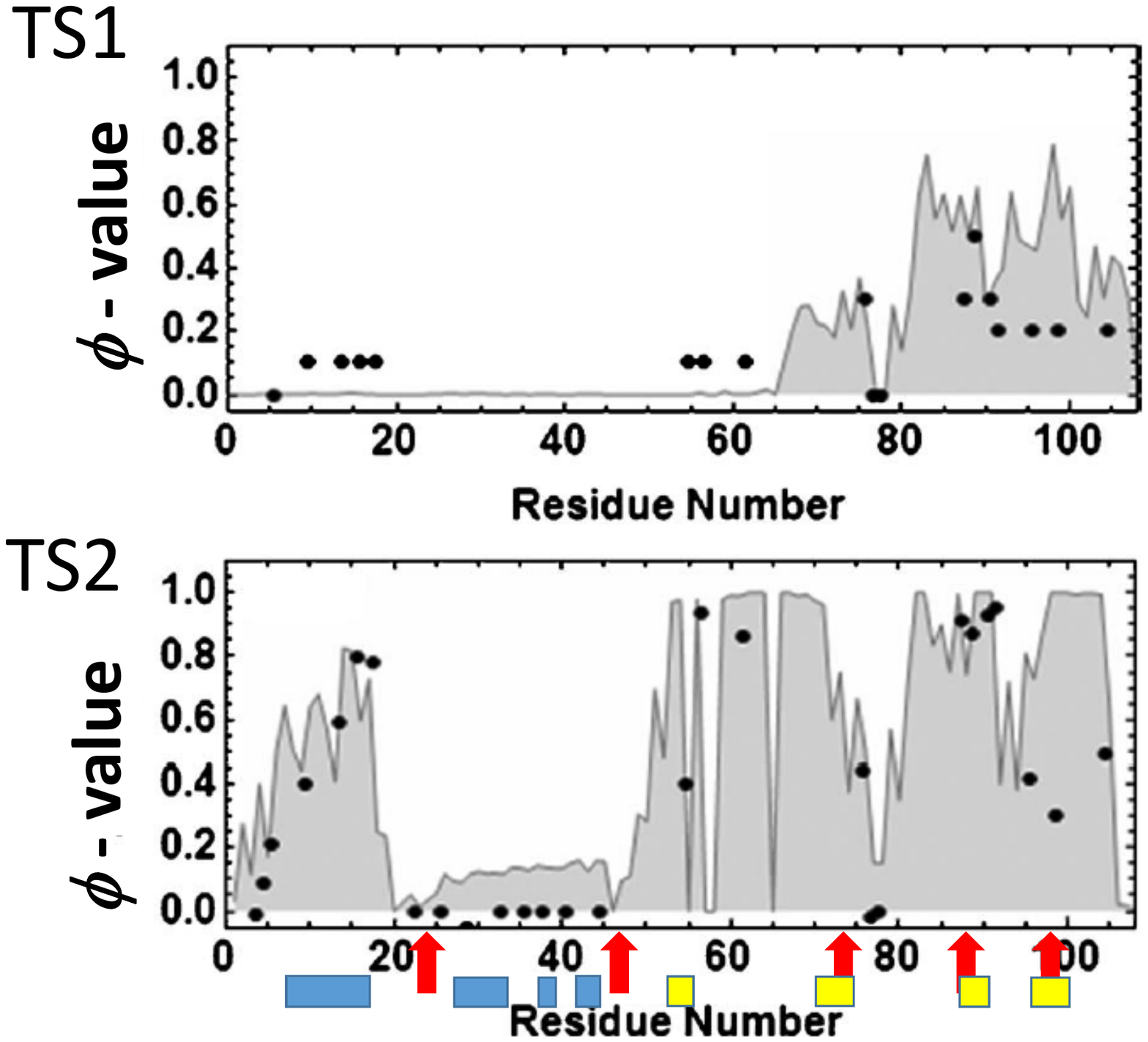}
\caption{
Calculated and observed $\phi$-values at the two transition states, TS1 and TS2, of barnase. Lines shaded with gray correspond to the calculated $\phi$-values with the WSME model. Dots are the experimentally observed values \cite{Serrano92, Salvatella05}. Red arrows are boundaries of modules defined by the pattern of atomic contacts in the native conformation \cite{Yanagawa93,Noguch93}. Bars shown on the bottom represent secondary structure elements, helices (blue) and strands (yellow). Modified from Fig.~15 of \cite{Itoh09} with permission.
}
\label{Fig:barnase_phi}
\end{figure}

Another example is shown for barnase in Figs.~\ref{Fig:barnase} and \ref{Fig:barnase_phi}. Barnase is a 110 residue $\alpha+\beta$ protein (Fig.~\ref{Fig:barnase}A), and its folding proceeds via an intermediate state \cite{Fersht99}. 
The two-dimensional  free-energy landscape $F(n_1,n_2)$ was calculated by disregarding two structurally unresolved residues with $N_1=54$ and $N_2=54$; therefore, $n_1$ is the order parameter of  folding for the N-terminal half and $n_2$ is the one for the C-terminal half.
In $F(n_1,n_2)$ of Fig.~\ref{Fig:barnase}B,  a dominant intermediate state is represented by a basin at a large $n_2$ and a small $n_1$ value, indicating that the C-terminal half is more structurally ordered than the N-terminal half is in the intermediate state. There are two transition states, TS1 between  the unfolded and intermediate states, and TS2 between the intermediate and native states.  In Fig.~\ref{Fig:barnase_phi}, the calculated $\phi$-values at TS1 and TS2 are compared with the experimentally observed values \cite{Serrano92, Salvatella05}, showing a good agreement between the WSME results and the observed data. In barnase,  as shown in Fig.~\ref{Fig:barnase_phi}, the $\phi$-value shows a large change around the boundaries of the structural modules, which are defined by the geometrical pattern of the native contacts  \cite{MGo87,MGo83, MGo81,Yanagawa93,Noguch93}. This interesting feature will be discussed later in the {\it Discussion} section.

As in the above examples, the WSME model explained the experimentally observed data of many proteins, which strongly suggests that the two major assumptions made in developing the WSME model, dominance of native interactions and the local cooperative formation of the native-like configuration, are indeed valid assumptions. 
The dominance of native interactions was also recently shown  \cite{Best13,Henry13} using folding trajectories of all-atom simulations performed by Shaw's group \cite{Larsen11,Piana12,Piana13}. Comparing the folding trajectories of  all-atom simulations and the WSME results, it was shown that the much simpler WSME model quantitatively explains the all-atom results \cite{Henry13}. The dominance of native interactions can be interpreted as following. 
When we consider the atomic details of a short molecular dynamics trajectory of the picosecond time-scale, there would be no distinction between native and non-native interactions; both have the same physical origin as electrostatic, hydrophobic, or van der Waals interactions. However, when we consider a microsecond or a longer process, the non-native interactions are only transiently formed within that process; also, the lifetime of native interactions is much longer due to the multi-residue cooperativity  forming the local ordered structure. Then, we can approximate the long-term process  using only the native interactions. The dominance of native interactions and the resulting globally biased energy landscape were first assumed by G$\bar{\rm o}$ and his colleagues to explain the two-state feature of folding transitions \cite{Go83, Taketomi75}. It was re-formulated later to explain how the trapping into the non-native states is prevented as well as how the Levinthal paradox is resolved in the energy landscape perspective \cite{Bryngelson95,Onuchic04}. Here, the dominance of native interactions in folding has been clearly supported by the results of the quantitative analyses of experimental data and all-atom simulations, and the WSME model has played an important role in these analyses.

By regarding the dominance of native interactions as the 0th order description,  non-native interactions should determine the next order description. Thus, non-native interactions should bring about the off-pathway intermediates in the folding process or work as ``friction'' in the course of folding \cite{Borgia12}; non-native interactions may destabilize the native conformation to some extent to make the structure flexible  to meet functional requirements \cite{Ferreiro14}. Understanding  the role of non-native interactions in long-term dynamics remains as an important  challenging problem.

In the WSME model, contiguous native-like segments are emphasized so that interactions such as those shown in Fig.~1B or 1C are neglected. Within a single-domain structure, this approximation seems reasonable. To make the native interaction between residues belonging to two segments separated by  residues with the non-native configuration effective, as shown in Fig.~1C, the multiple intervening residues in the linker between two segments must follow multiple non-native directions to  compensate for ``incorrect'' directions and  to recover the ``correct'' orientation between  residues having the native interaction. This flexible structural adjustment of the linker chain is a necessary condition to make the interaction effective, but such flexible adjustment is rare in a single domain when the linker is short. Therefore, the assumption made for the WSME model is considered appropriate  at least for describing the folding process of single-domain proteins.  Indeed, the validity of the WSME model was shown for single-domain proteins \cite{Henry04, Itoh06, Kubelka08, Nelson08, Yu08, Itoh09, Henry13}, but further careful argument is necessary to describe  multi-domain proteins, particularly when they have a nontrivial topological arrangement of domains, as discussed in the next section.

\section*{The eWSME Model for Multi-domain Proteins}

\begin{figure}[htbp]
\includegraphics[width=8cm]{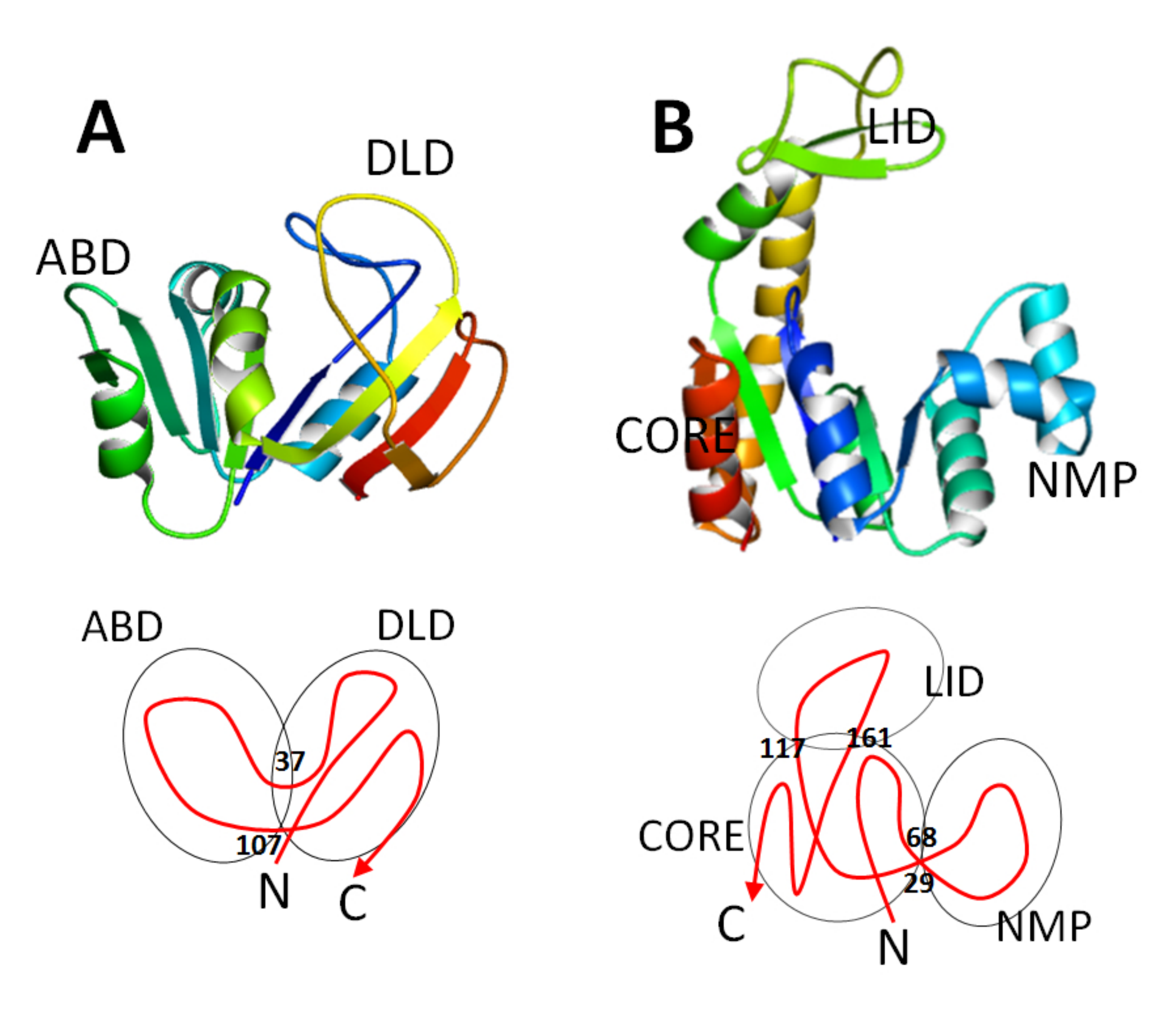}
\caption{
Examples of multi-domain proteins with non-trivial topology. A) Dihydrofolate reductase (DHFR) (PDB code:~1rx1) has two domains, DLD and ABD. B) Adenylate kinase (AdK) (PDB code:~4ake) has three domains, CORE, NMP, and LID. Topological connectivity of the chain is illustrated at the bottom.
}
\label{Fig:topology}
\end{figure}

Many proteins show all-or-none two-state transitions between the folded and unfolded states, but in 1978,  Wako and Sait$\hat{\rm o}$ \cite{Wako78b} suggested the presence of an intermediate state for lysozyme  based on the calculated heterogeneous size distribution of contiguous native-like segments.  In the 1980s,  clear experimental evidence was discovered for the folding intermediates, which were referred to as the molten globule states \cite{Arai00}. Particularly, the folding process of typical small multi-domain proteins, such as $\alpha$-lactalbumin and lysozyme, was analyzed. It was shown that, in these example proteins, the intermediate state in the equilibrium three-state transition is very similar to the intermediate state that appears on the kinetic folding pathway, suggesting the pivotal role of the molten globule state in protein folding. Furthermore, the structure of the molten globule state is heterogeneous and  composed of ordered and disordered parts, whereas the degrees of compaction and side-chain packing  largely depend on the protein species. To obtain a unified picture of the diversity of the molten globule state, extending the WSME model to  describe generic multi-domain proteins by taking account of native interactions, as illustrated in Fig.~1C, is strongly desired. The need for considering native interactions between residues separated by others with non-native configuration is evident particularly for  proteins having topologically complex structures, as shown in Fig.~\ref{Fig:topology}.

Dihydrofolate reductase (DHFR), a 159 residue $\alpha$/$\beta$ protein, for example,  has two domains, the discontinuous loop domain (DLD) and the adenosine-binding domain (ABD), as shown in Fig.~\ref{Fig:topology}A; the ABD is a continuous domain comprising the residues 38-106, and the DLD is a discontinuous domain comprising the N-terminal part (residues 1-37) and the C-terminal part (residues 107-159). Therefore, native interactions between the N- and C-terminal parts in the  DLD are expected to form even when the intervening ABD is disordered, which is just the case illustrated in Fig.~1C. A convenient way to consider such interactions is to introduce a virtual link connecting the N- and C-termini  (Fig.~1D) and applying the WSME Hamiltonian to this virtually closed ring  to derive the partition function $Z_{\rm ring}$. Using  $Z_{\rm ring}$,  the extended WSME (eWSME) partition function is defined by
\begin{eqnarray}
Z_{\rm eWSME}(n)=Z_{\rm WSME}(n)+(Z_{\rm ring}(n)-Z_{\rm WSME}(n))e^{ S_{\rm ring}(n)/k_{\rm B} },
\end{eqnarray}
where $S_{\rm ring}(n)<0$ is the entropic reduction arising from the constraint to place the N- and C-termini at a distance determined by the native conformation, which can be estimated  assuming that the disordered parts of the chain under the $n$ constraint  behave as fragments with random configurations \cite{Inanami14}. $Z_{\rm eWSME}$ is smoothly interpolated between $Z_{\rm WSME}$ and $Z_{\rm ring}$; $Z_{\rm eWSME}\approx Z_{\rm WSME}$, when the entropic reduction is significant, as $S_{\rm ring}\ll 0$, and  $Z_{\rm eWSME}\approx Z_{\rm ring}$, when the entropic reduction is negligible, as $S_{\rm ring}\approx 0$. $Z_{\rm eWSME}$ incorporates both local multi-residue correlations as in $Z_{\rm WSME}$ and  native interactions separated by intervening non-native residues with suitable statistical weights; also, it is exactly calculable.

\begin{figure}[htbp]
\includegraphics[width=12cm]{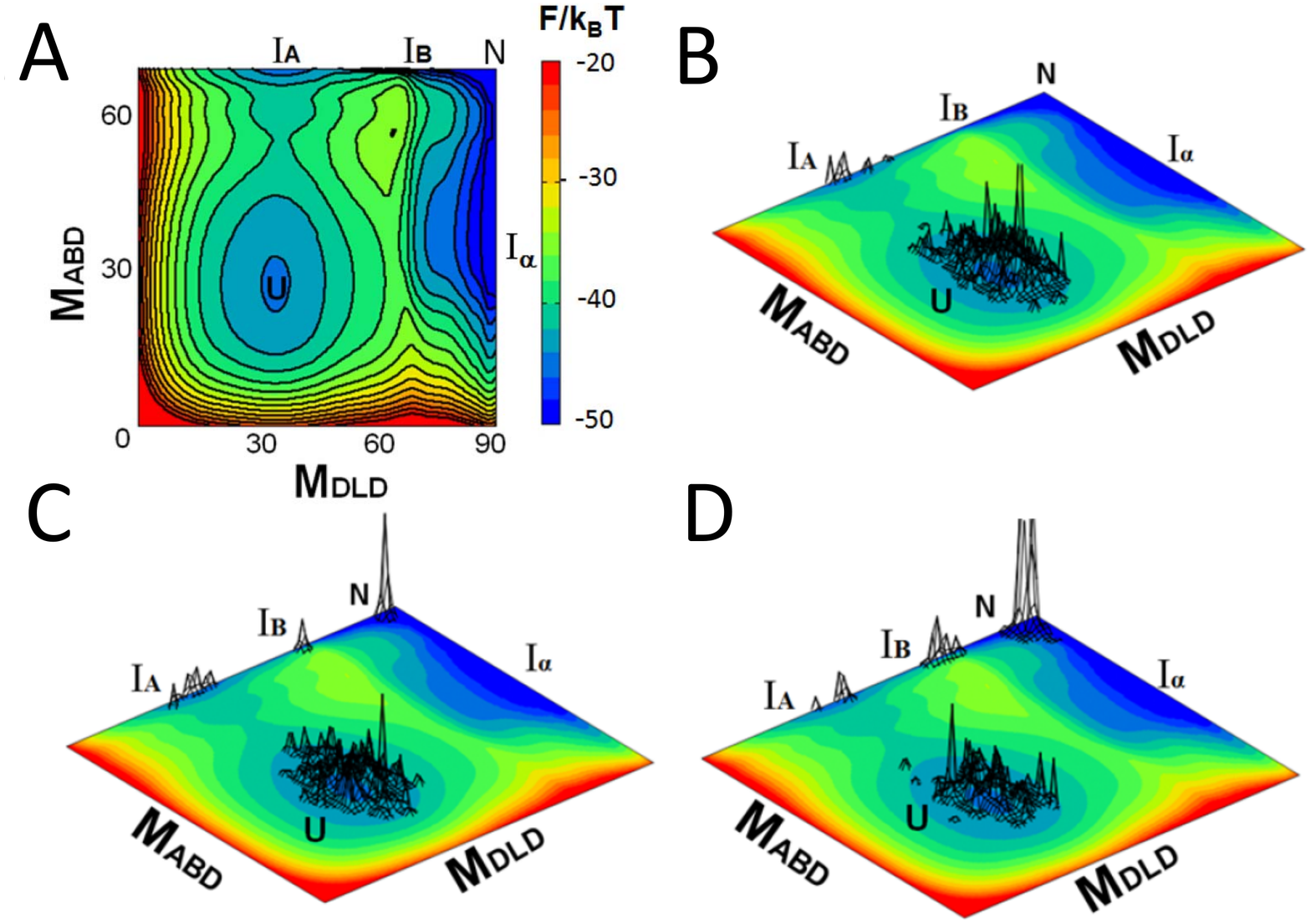}
\caption{
Free-energy landscape and kinetics of DHFR folding calculated by the eWSME model. A) Free-energy landscape of DHFR folding  represented in the two-dimensional space of $M_{\rm DLD}$ and $M_{\rm ABD}$. The landscape has basins corresponding to the unfolded state U, the native state N, and the intermediates, I$_{\rm A}$, I$_{\rm B}$, and I$_\alpha$. B--D) Evolution of the population of 200 molecules simulated with the Monte Carlo calculation at B) $3.3 \times 10^5 t_0$, C) $1.6 \times 10^6 t_0$, and D) $3.0 \times 10^6 t_0$, where $t_0$ is a unit of time in simulation. Reproduced from \cite{Inanami14}.
}
\label{fig:DHFR_landscape}
\end{figure}

The two-dimensional free-energy folding landscape of DHFR calculated with this eWSME model is shown in Fig.\,\ref{fig:DHFR_landscape}A \cite{Inanami14}. Here, the two-dimensional space is defined by the  parameters $M_{\rm DLD}=\sum_{i\in {\rm DLD}}m_i$ and $M_{\rm ABD}=\sum_{i\in {\rm ABD}}m_i$. This landscape has basins at $(M_{\rm DLD},M_{\rm ABD})\approx (30,30)$, which is the basin of the unfolded state (U); at $(M_{\rm DLD},M_{\rm ABD})\approx (30,69)$ (the basin denoted by I$_{\rm A}$); at $(M_{\rm DLD},M_{\rm ABD})\approx (70,69)$ (the basin I$_{\rm B}$); at $(M_{\rm DLD},M_{\rm ABD})\approx (90,35)$ (the basin I$_\alpha$); and at $(M_{\rm DLD},M_{\rm ABD})\approx (90,69)$ (the basin of the native state, N).  In I$_{\rm A}$,  the ABD is folded and the  DLD is unfolded, whereas, in I$_\alpha$, the DLD is folded and the ABD is unfolded.
The basin I$_\alpha$ has lower free energy than I$_{\rm A}$; however, I$_\alpha$ is separated from U by a higher free-energy barrier than I$_{\rm A}$. Therefore, we can expect that molecules starting from U pass through I$_{\rm A}$ to proceed along the pathway ${\rm U}\rightarrow {\rm I}_{\rm A}\rightarrow {\rm I}_{\rm B}\rightarrow {\rm N}$. This was confirmed by numerically following the kinetic change of $\{ m_i\}$ with the Monte Carlo simulation using the following function to calculate the effective eWSME energy for the Metropolis criterion;
\newpage
\begin{eqnarray}
E_{\rm eWSME} (\{m_i\})&=& \nonumber \\
&-&k_{\rm B}T \log \left (
 e^{-H_{\rm WSME}/k_{\rm B}T} + (e^{-H_{\rm ring}/k_{\rm B}T}- e^{-H_{\rm WSME}/k_{\rm B}T} )e^{ S_{\rm ring}/k_{\rm B}}
\right ) \nonumber \\
&+& k_{\rm B}T\sum_i \sigma_i m_i.  \nonumber \\
\end{eqnarray}
The kinetic evolution of the DHFR molecules' population on the two-dimensional space is shown in Figs.\,\ref{fig:DHFR_landscape}B--\ref{fig:DHFR_landscape}D. These panels show that the population indeed proceeds along the folding pathway ${\rm U}\rightarrow {\rm I}_{\rm A}\rightarrow {\rm I}_{\rm B}\rightarrow {\rm N}$ by sequentially visiting the intermediate states I$_{\rm A}$ and I$_{\rm B}$. This pathway agrees with the observed pathway and kinetics of folding \cite{Arai2011}. This sequential pathway is preferred due to the high free-energy barrier between U and I$_\alpha$, which prevents folding trajectories from branching to I$_\alpha$. This barrier arises from the large entropy decrease, which brings together the discontinuous parts to form DLD. In other words, the topological complexity of DHFR is the reason for this simple sequential pathway of folding. 
It should also be noted that the free-energy barrier between N and I$_\alpha$ is predicted to be low, leading to structural fluctuations, including the partial unfolding/folding of the ABD that can be important for the function of DHFR in the native state.

\begin{figure}[htbp]
\includegraphics[width=12cm]{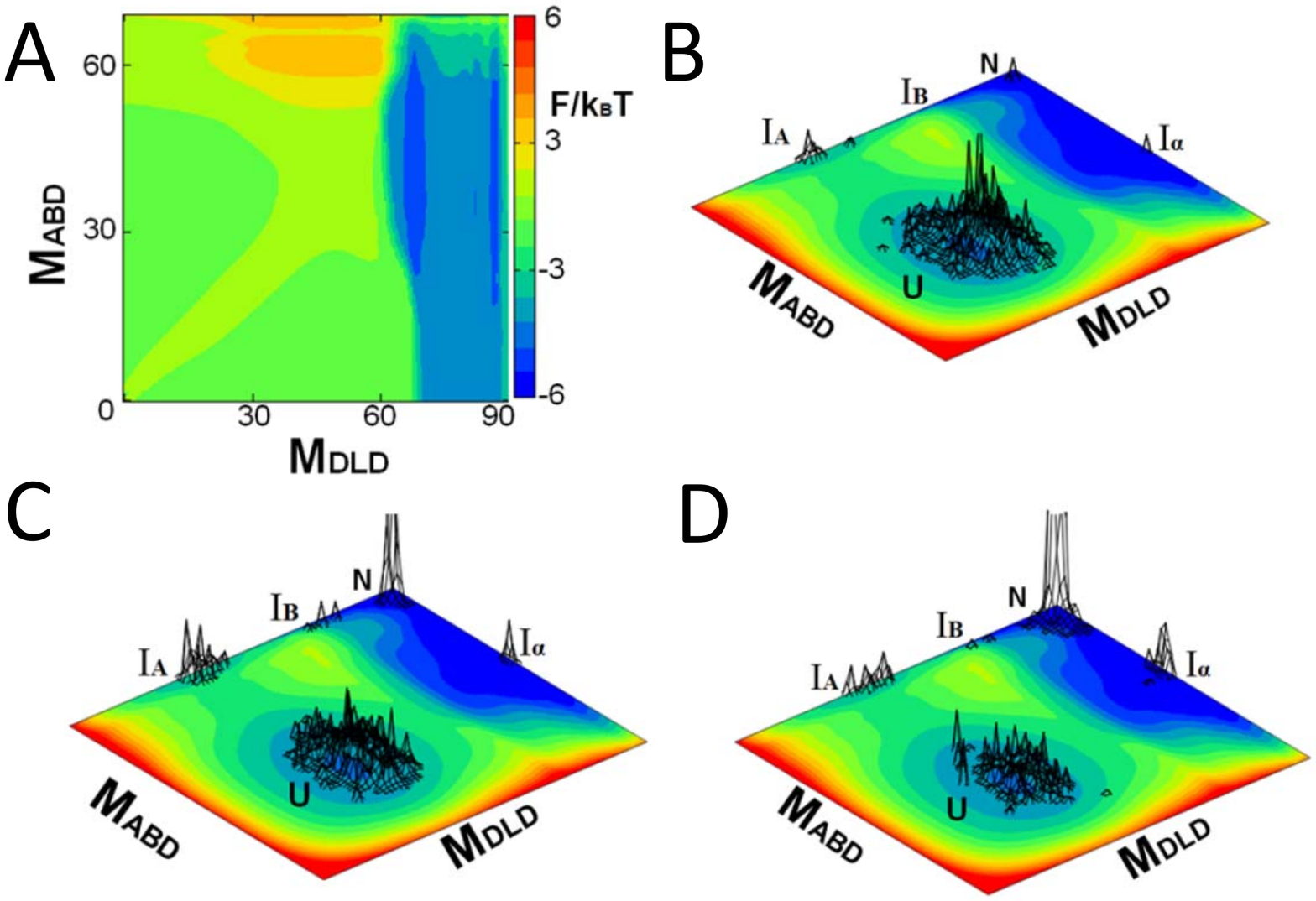}
\caption{
Free-energy landscape and folding kinetics of the circular permutant of DHFR calculated by the eWSME model. A) Difference in the free-energy landscape between the wild type and the circular permutant of DHFR. B--D) Evolution of the population of 200 molecules simulated with the Monte Carlo calculation at B) $3.3 \times 10^5 t_0$, C) $1.6 \times 10^6 t_0$, and D) $3.0 \times 10^6 t_0$, where $t_0$ is a unit of time in  simulation. Reproduced from \cite{Inanami14}.
}
\label{fig:DHFR_permutation}
\end{figure}

We should note that the topological complexity of DHFR can be resolved by circular permutation. Connecting the N and C termini and disconnecting the linker part of the chain between DLD and ABD, both ABD and  DLD become continuous domains comprising continuous parts of the chain. The free energy change due to this circular permutation was calculated by the eWSME model and shown in Fig.\,\ref{fig:DHFR_permutation}. This circular permutation increases the free energy at around I$_{\rm B}$ and lowers the free energy at the barrier between U and I$_\alpha$. Then, the kinetic evolution of  DHFR molecules' population branches into two pathways, ${\rm U}\rightarrow {\rm I}_{\rm A}\rightarrow {\rm I}_{\rm B}\rightarrow {\rm N}$ and ${\rm U}\rightarrow {\rm I}_{\alpha}\rightarrow {\rm N}$, as indicated by the Monte Carlo results of Figs.\,\ref{fig:DHFR_permutation}B--\,\ref{fig:DHFR_permutation}D. In this way, the simplification of the DHFR topology through circular permutation brings about the complex folding behavior. This complex folding behavior is consistent with the observed  folding kinetics of the circular permutant \cite{Texter1992}.

Further extension of the WSME model is possible for proteins with more complex topologies, and we here outline  this idea. Adenylate kinase (AdK), for example, has three domains: CORE (residues 1-29, 68-117, and 161-214),  NMP (residues 30-67), and LID (residues 118-167), as shown in Fig.~\ref{Fig:topology}B.
We define the virtual ring closures at residues 29 and 68 (closure-1), 117 and 161 (closure-2), and  1 and 214 (closure-3). The WSME partition function $Z_{\rm ring}(i)$ is calculated by assuming only one closure for $i=$1, 2, or 3,   $Z_{\rm ring}(ij)$  is calculated for two closures with $ij=$12, 23, or 31, and $Z_{\rm ring}(123)$ is calculated for three closures. Then, $Z_{\rm eWSME}$ is calculable from the WSME Hamiltonian as
\begin{eqnarray}
Z_{\rm eWSME}&=&Z_{\rm WSME}  
+ (Z_{\rm ring}(1)-Z_{\rm WSME}) A_1(1-A_2)(1-A_3) \\ \nonumber
&+& (Z_{\rm ring}(2)-Z_{\rm WSME}) A_2(1-A_3)(1-A_1) \\ \nonumber
&+& (Z_{\rm ring}(3)-Z_{\rm WSME}) A_3(1-A_1)(1-A_2) \\ \nonumber
&+& (Z_{\rm ring}(12)-Z_{\rm WSME}) A_1A_2(1-A_3) + (Z_{\rm ring}(23)-Z_{\rm WSME}) A_2A_3(1-A_1) \\ \nonumber
&+& (Z_{\rm ring}(31)-Z_{\rm WSME})A_3A_1(1-A_2) + (Z_{\rm ring}(123)-Z_{\rm WSME}) A_1A_2A_3,
\end{eqnarray}
where $A_i=\exp(S_{\rm ring}(i)/k_{\rm B})$ is a factor representing the entropy reduction due to the closure-$i$, 
which could be estimated by evaluating the probability that the two sites in a Gaussian chain are located at the closure distance from each other, under the constraint of a given pattern of $\{ m_i\}$.
In this way,  the eWSME model can be directly applied to proteins with various topologies, as exploring folding mechanisms of multi-domain proteins with a unified perspective is an important avenue of the folding studies.

\section*{The aWSME Model for Protein Allostery}
The classical view of protein folding, wherein folding proceeds along a definite pathway \cite{Baldwin95}, was replaced by the modern energy landscape picture, which describes protein folding as fluctuating diffusive motions over a globally biased energy landscape. Energy landscape methods have shown that the folding pathway and transition state ensemble are determined by the statistical features of the distributed fluctuating trajectories; these methods enabled the quantitative understanding of protein folding and guided methods of protein engineering \cite{Onuchic04}. The energy landscape perspective should be important not only for protein folding but also for protein conformational change, wherein fluctuations and diversity of trajectories are significant. Particularly, the energy landscape description should be necessary for understanding allosteric transitions \cite{Boehr06, Terada13, Tsai14}.

An allosteric transition is a change in the distribution of a protein's structure triggered by a chemical or physical perturbation \cite{Motlagh14}, which is often an essential step for  proteins to exert their functions. Although the classical view of allosteric transition is based on the picture of a  deterministic sequential structural change \cite{Vreede10}, motions in allosteric transition should  bear flexible stochastic fluctuations that may  allow diversely different transition trajectories, as in protein folding, which should be  quantitatively assessed by energy landscape methods. 
For this purpose, the WSME model can be extended to describe the energy landscape of allosteric transitions.

Here, we assume that a protein shows two different low-energy conformations in the native state. To be more specific, we consider the case that one  is the  active (A) conformation, which has the higher affinity to bind a partner protein, and the other is the inactive (I) conformation, which has the lower affinity to bind it. The dominant conformation, around which the protein structure fluctuates, switches from I to A upon binding of a ligand or through  chemical modification such as phosphorylation of the protein.  We should note that the following theoretical scheme is applicable to cases other than this I-A structural change when the transition between two low-energy conformations is concerned with.
We assume that $m_i$ can take three values, A, I, and D; $m_i={\rm A}$ or I when the $i$th residue takes the configuration similar to that found in the A or I conformation, respectively, and  $m_i={\rm D}$, when the residue takes a disordered non-native configuration. Here, for mathematical convenience, to calculate the partition function from the Hamiltonian, we use a redundant expression of either  $m_i={\rm A}$ or $m_i={\rm I}$ for the residue with the configuration common to A and I \cite{Itoh10}.

The contact patterns in the native conformations are expressed as $\Delta^{\rm A}_{ij}$ and $\Delta^{\rm I}_{ij}$; $\Delta^{\rm A(or\, I)}_{ij}=1$ when the residues $i$ and $j$ are in contact in the A(or I) conformation and $\Delta^{\rm A(or\, I)}_{ij}=0$, otherwise.  $\Delta^{\rm C}_{ij}=\Delta^{\rm A}_{ij}\Delta^{\rm I}_{ij}$ represents the contact pattern which is common to A and I. ${\widetilde{\Delta}}^A_{ij}=\Delta^A_{ij}(1-\Delta^C_{ij})$ and ${\widetilde{\Delta}}^{\rm I}_{ij}=\Delta^{\rm I}_{ij}(1-\Delta^{\rm C}_{ij})$ are the contact patterns which are specific to A and I, respectively. 
We define the functions $P^{\rm A}_k(m_k)$, $P^{\rm I}_k(m_k)$, and $P^0_k(m_k)$ by $P^{\rm A}_k({\rm A})=1$, $P^{\rm A}_k({\rm I})=P^{\rm A}_k({\rm D})=0$, $P^{\rm I}_k({\rm I})=1$, $P^{\rm I}_k({\rm A})=P^{\rm I}_k({\rm D})=0$, and $P^0_k(m_k)=P^{\rm A}_k(m_k)+P^{\rm I}_k(m_k)$. Then, the WSME Hamiltonian for allosteric transition (the aWSME Hamiltonian) is
\begin{eqnarray}
H_{\rm aWSME}(\alpha,\{m_i\})&=&V_{\alpha}(\{m_i\})  \nonumber \\
&+&\sum_{i=1}^{N-1}\sum_{j=i+1}^{N}\epsilon_{ij}\left( 
\Delta^{\rm C}_{ij}\prod_{k=i}^j P^0_k(m_k)+{\widetilde{\Delta}}^{\rm A}_{ij}\prod_{k=i}^j P^{\rm A}_k(m_k)+{\widetilde{\Delta}}^{\rm I}_{ij}\prod_{k=i}^j P^{\rm I}_k(m_k)\right), \nonumber \\
\end{eqnarray}
where  $\alpha$ distinguishes the ligand binding/unbinding or the phosphorylation/dephosphorylation and $V_{\alpha}(\{m_i\})$ represents the local interactions between the bound ligand and surrounding residues or those around the phosphorylated site \cite{Itoh10}. 
The first term in the summation of the right-hand side of Eq.~6 is the energy decrease due to the many-residue correlation to form native-like segments, and the second and third terms represent the energy decrease due to the many-residue correlation to form A and I-like segments, respectively.
We define the order parameter $n$ of the folding and the order parameter $x$ of allostery as $n=\sum_{i=1}^N P^0_k(m_i)/N$ and $x=M_{\rm A}/N_{\rm A}$, respectively. Here, $M_{\rm A}$ is the number of residues assuming the configuration specific to the A conformation, and $N_{\rm A}$ is the maximal number of $M_A$, so that $(x,n)=(0,1)$ is the I conformation, $(x,n)=(1,1)$ is the A conformation, and $(x,n)=(0,0)$ is the completely disordered state. The partition function $Z_{\rm aWSME}(\alpha,x,n)$ and the  two-dimensional free-energy landscape $F_{\alpha}(x,n)=-k_{\rm B}T\log Z_{\rm aWSME}$ are exactly calculable from $H_{\rm aWSME}$. See \cite{Itoh11} for a more detailed explanation of the model.

\begin{figure}[htbp]
\includegraphics[width=8cm]{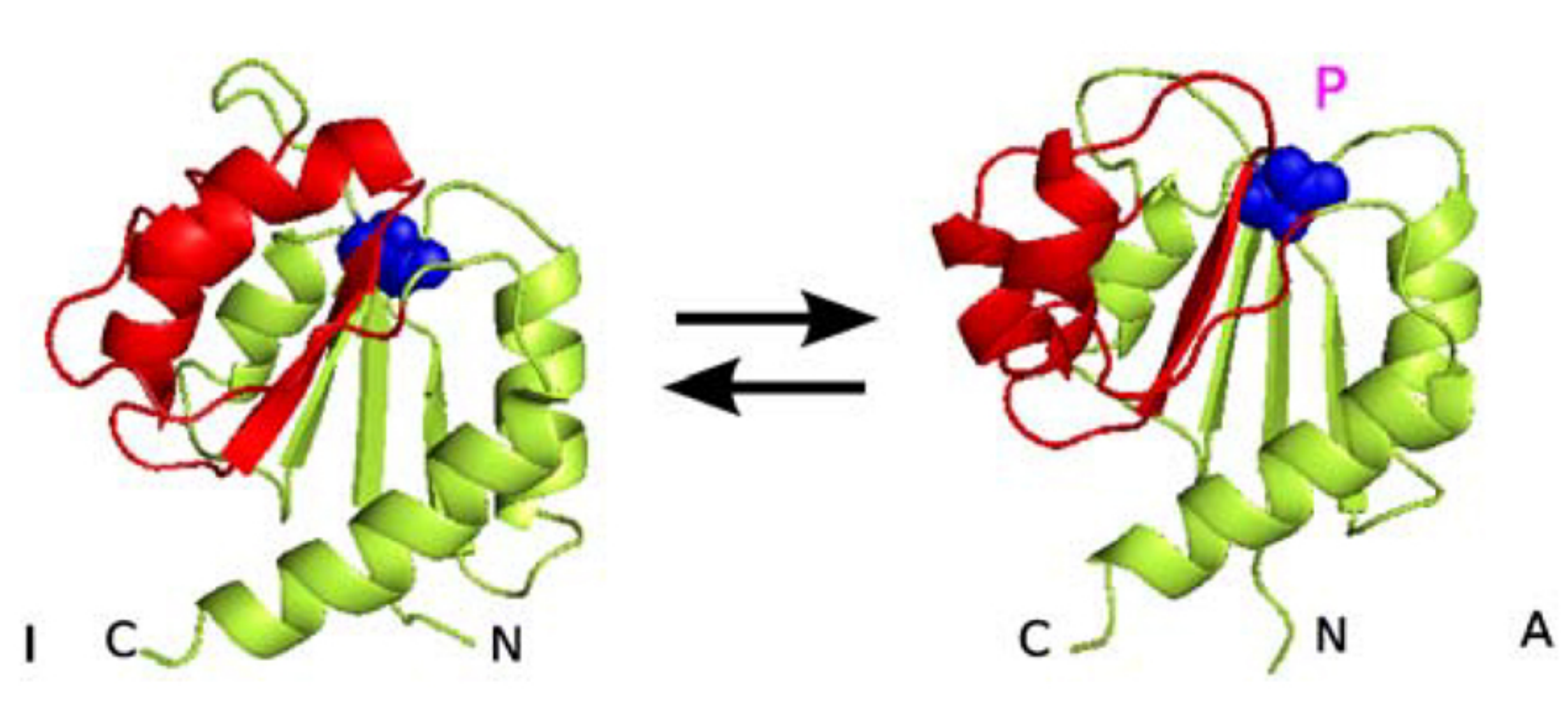}
\caption{
Allosteric transition of NtrC. Upon phosphorylation of Asp54, the NtrC structure switches from a state around the inactive  (I) conformation (PDB code:~1dc7) to another state around the active (A) conformation (PDB code:~1dc8). Asp54 is shown with blue colored spheres. ``3445 face'' (the region comprises helices and strands, $\alpha$3, $\beta$4, $\alpha$4, and $\beta$5) is colored red.  Reproduced from \cite{Itoh10}.
}
\label{Fig:NtrC}
\end{figure}

\begin{figure}[htbp]
\includegraphics[width=10cm]{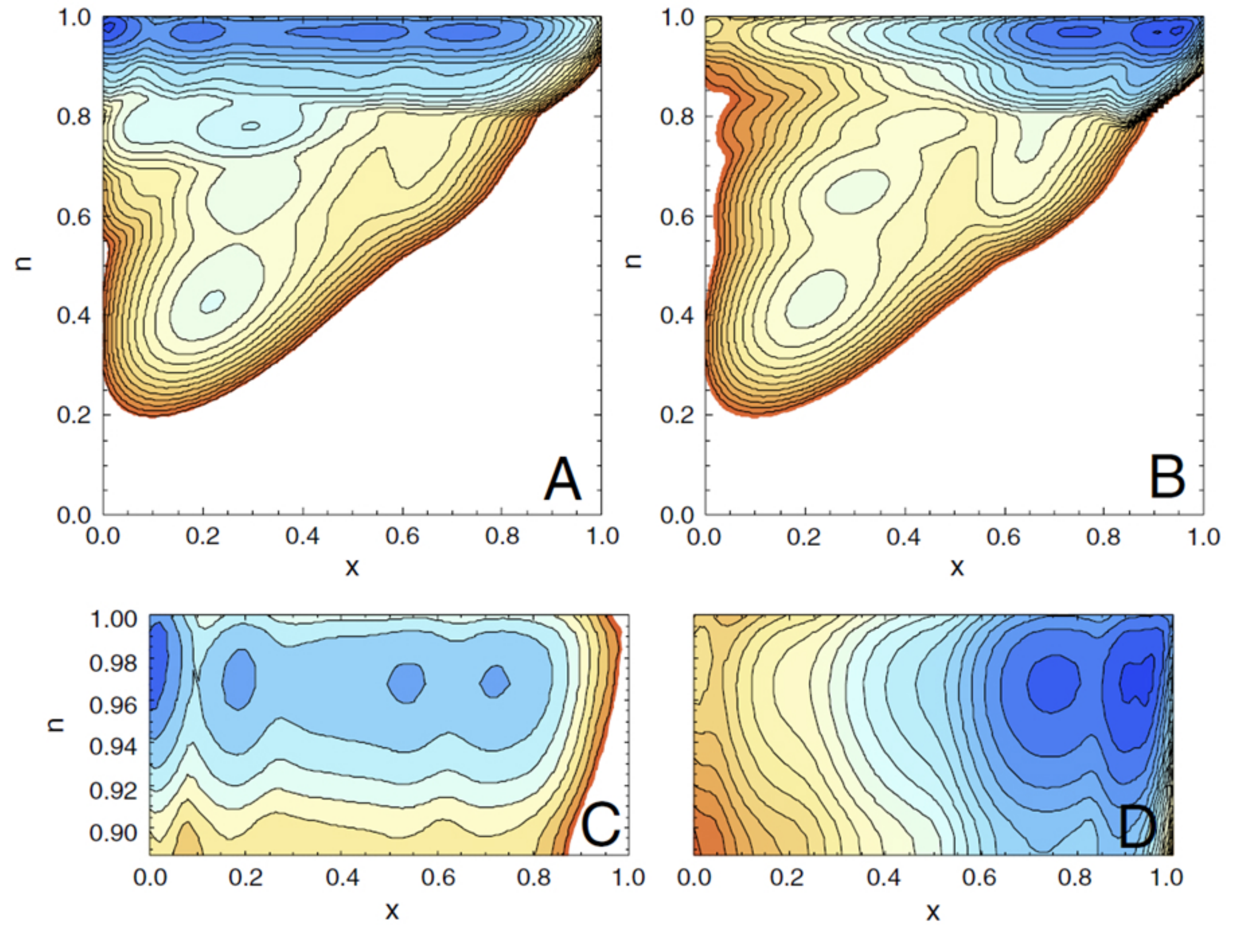}
\caption{
Free-energy landscape $F_{\alpha}(x,n)$ of allosteric transition of NtrC calculated with the aWSME model. $x$ is the order parameter of allosteric transition and $n$ is the order parameter of folding transition. $(x,n)=(0,1)$ is the I conformation, $(1,1)$ is the A conformation, and $(0,0)$ is the completely disordered state. A) $F_{\rm dephos}(x,n)$ in the dephosphorylated state and B) $F_{\rm phos}(x,n)$ in the phosphorylated state. C) and D) are closeups of A) and B), respectively, at $n\approx 1$. Contour is drawn for every $2k_{\rm B}T$. Reproduced from \cite{Itoh10}.
}
\label{Fig:NtrC_landscape}
\end{figure}

\begin{figure}[htbp]
\includegraphics[width=10cm]{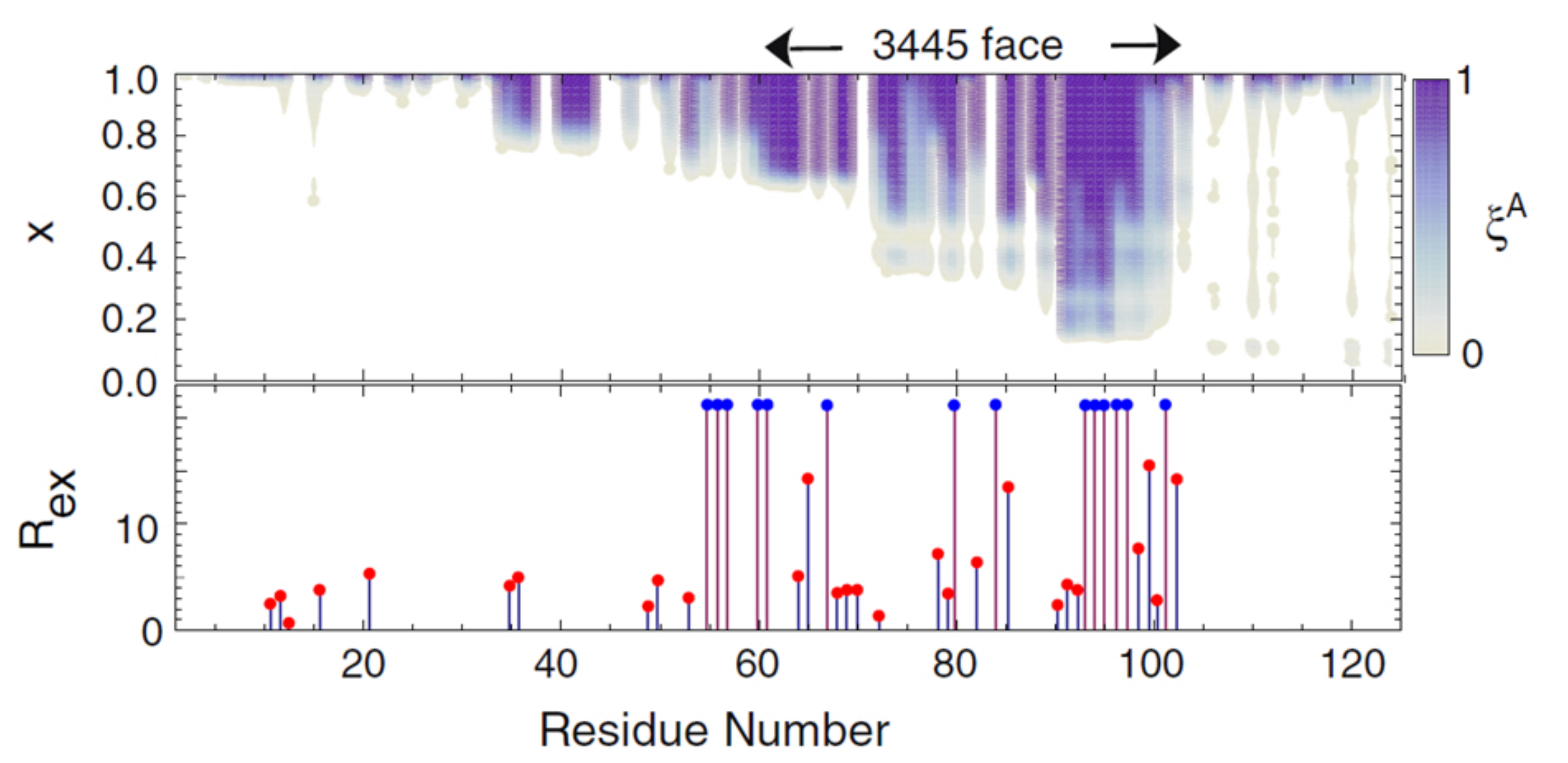}
\caption{
Pre-existing structural fluctuation of NtrC. (Top) The parameter $\xi^{\rm A}$ showing the extent of the A-like structure development in the dephosphoryated state. $\xi^{\rm A}$ calculated with the aWSME model under the constraint of each fixed $x$ and $n=1$ is plotted in gray scale. Even in conformations near the I conformation with small $x$, the A-like structure appears as a fluctuation around the 3445 face. (Bottom) $R_{\rm ex}$ observed in the relaxation measurement of NMR in the dephosphorylated state \cite{Volkman01} are shown with red dots.  $R_{\rm ex}$ is larger than a threshold  for the blue dots \cite{Volkman01}. Reproduced from \cite{Itoh10}.
}
\label{Fig:NtrC_Rex}
\end{figure}

Fig.~\ref{Fig:NtrC} illustrates the allosteric transition of an example protein, the bacterial nitrogen regulatory protein C (NtrC). The distribution of the NtrC structures is dominated by the A conformation, when  the residue Asp54 is phosphorylated, and by the I conformation, when dephosphorylated.  Fig.~\ref{Fig:NtrC_landscape} shows $F_{\alpha}(x,n)$ calculated with the aWSME model. Although the most stable structure in $F_{\rm dephos}(x,n)$ is the I conformation at  $(x,n)\approx (0,1)$, a low free-energy valley extends from I to  A conformations with metastable basins at $(x,n)\approx (0.2, 0.97)$, (0.55, 0.97), and (0.75, 0.97), demonstrating that the dephosphorylated NtrC should exhibit large structural fluctuation. The NtrC molecules within the valley bear the A-like features, which transiently appear as fluctuations, though the most stable structure is the I conformation. 
As shown in Fig.~\ref{Fig:NtrC_Rex}, this structure fluctuation explains the observed $R_{\rm ex}$ values derived from  the $R_1$, $R_2$, and the NOE relaxation  data of NMR \cite{Volkman01}.

As shown in Fig.~\ref{Fig:NtrC_landscape}, when Asp54 is phosphorylated, a basin that does not exist in $F_{\rm dephos}(x,n)$ appears at $(x,n)=(0.95,0.97)$ in $F_{\rm phos}(x,n)$. Therefore, the conformation  close to A becomes most stable upon phosphorylation. The large fluctuation between A and I in the dephosphorylated state shows that the transition  from I to A can be regarded as the selection of pre-existing A-like conformations, but the shift from (0.75, 0.97) to (0.95, 0.97) shows that the ``induced-fit''   works during the last step of this transition. Thus, the aWSME model reveals that the mixed mechanisms of conformation selection and induced fit regulate the allosteric transition of NtrC.

The large structural fluctuation in the dephosphorylated state is due to the entropic gain for the intermediate $x$. In the intermediate $x$ regime, multiple A- or I-like segments coexist in the chain, and a large number of mosaic patterns of these segments are possible; this large number of structures is the reason for the large entropy in this regime. In other words, the multitude of fluctuating trajectories with similar energies is the reason for the flat free-energy landscape and  large fluctuation along the $x$ variance with $n\approx 1$. Such entropic gain is not taken into account by conventional simulations based on the classical picture assuming a unique definite transition pathway. Thus, the results of the WSME model reveal the importance of fluctuating movement over the energy landscape. It should be noted that in the problem of allostery, the landscape itself is modified by binding/unbinding of an effector such as the phosphate group, inducing the dynamical transition $F_{\rm dephos}\leftrightarrow F_{\rm phos}$. To emphasize this aspect, we would argue that  the ``dynamical energy landscape  view'' is important for analyzing protein allostery and functions.

Finally, we note that the aWSME model can be applied to the folding problem, when  competition between the native conformation and an off-pathway intermediate state with a distinct non-native structure dominates the folding process \cite{Hamada96,Borgia15}. The aWSME model is applicable to this problem using these native and non-native conformations in place of the A and I conformations in the above analysis.

\section*{Discussion: Cooperativity and Modularity}
Prof. Nobuhiko Sait$\hat{\rm o}$ emphasized the importance of the hierarchal pathway of protein folding through the WSME model development and the related models of secondary structure formation  \cite{Wako83,Saito88,Saito99}. In this hierarchical picture, ``islands'' or local native-like contiguous segments are spontaneously formed at the early stage of folding, and  folding proceeds through growth and coalescence of these segments through long-range interactions. Sait$\hat{\rm o}$ suggested that the segments  formed first should typically be secondary structure elements (SSEs), such as $\alpha$-helices or $\beta$-strands, and these SSEs are packed with hydrophobic interactions in the later stage of folding \cite{Wako83,Saito88,Saito99}. However,  in many cases, the loop regions include as dense hydrogen-bonds or other interactions as in SSEs such that local structures including  loops can be  energetically stabilized similarly to SSEs. Therefore, segments that include loops could also be formed during the early stage of folding. A well-known example of a loop, where the folding reaction initiates, is the distal  hairpin loop of src SH3 \cite{Grantcharova98}. The above discussion suggests that  we should carefully examine the parts of the protein that fold during the early stage of the folding process.  Importantly, the statistical weight of the different folding  pathways can be  compared with the WSME model by taking account the  balance between energy and entropy so that the quantitative comparison between the experiments and the WSME results would facilitate  solving this problem.

Local segments, which could be identified as units of a protein's substructure, have been defined and analyzed from several viewpoints. A notable approach is the geometrical analysis;  using the contact pattern in the native conformation, ``modules'' were defined as units of the substructure \cite{MGo83}.  G$\bar{\rm o}$ showed that the boundaries of these modules coincide with the boundaries of exons of example proteins \cite{MGo87,MGo81}, which suggested  that modern proteins were formed through shuffling of modules in the evolutionary history. Barnase, for example, comprises six modules, M1, M2, $\cdots$, M6, and their boundaries are at residues 24, 52, 73, 88, and 98 \cite{Yanagawa93,Noguch93}. In Fig.~\ref{Fig:barnase_phi}, these module boundaries are compared with the calculated and observed $\phi$-values at two transition states, TS1 and TS2. Meanwhile, when we examine an ensemble of numerous protein molecules, those molecules diffusively move  on the energy landscape to diversely trace different trajectories so that the transition state, in which the folding nucleus is formed, is not dominated by a  unique structure, but should be described as an ensemble of many heterogeneous structures. The $\phi$-values represent the average tendency to form the ordered structure at each residue in this transition state ensemble.

We found distinct dips in the calculated $\phi$-values at residues 72--73 and 89--90 at TS1, and at  20--23, 46, 72--73, 77--78, and 87--89 at TS2, showing the rough correlation between the module boundaries and the $\phi$-value boundaries. Through this comparison, we see that in the nucleus formation in  TS1, M1 (residues 1--24) and M2 (residues 25--52) are disordered, M3  (residues 53--73) and M6 (residues 99--110) have small but finite probability of structure formation, and M4 (residues 74--88) and M5(residues 89--98) have intermediate levels of probability of folding. In another stage of nucleus formation in TS2, M1 has an intermediate level of probability of folding, M2 is disordered, and M3--M6 have  higher probabilities of folding.
Although the correspondence is not exact, this comparison suggests that module-like segments are formed at the transition states of barnase as cooperative structure formation units.

Energetic analysis is another method to define the subunits. Using a knowledge-based potential, the units of cooperative folding, foldons, were defined as segments that show the maximal energy gap between ordered and disordered structures \cite{Panchenko96,Panchenko97}. For barnase, the foldons' boundaries do not exactly match  with those of the modules; however, there is a correlation between them; foldon-1 corresponds to M1, foldon-2 corresponds to M2, and foldon-3 corresponds to a part extending from M3 to M6  \cite{Panchenko96}. With this terminology, foldon-3 is folded with a large probability, foldon-1 is folded with a modest probability, and foldon-2 is almost unfolded at TS2 of barnase.

Comparing multiple proteins showed that there are correlations among modules, exons, and foldons, but the correspondence is not perfect and  deviations specific to proteins were reported \cite{Panchenko96,Panchenko97}. To elucidate the correlation and deviation of these differently defined local segments, the comprehensive  comparison of different types of proteins is necessary.  As shown in the above discussion,  the $\phi$-value analysis with the WSME model should be useful for interpreting the results of such a comparison.

\begin{figure}[htbp]
\includegraphics[width=10cm]{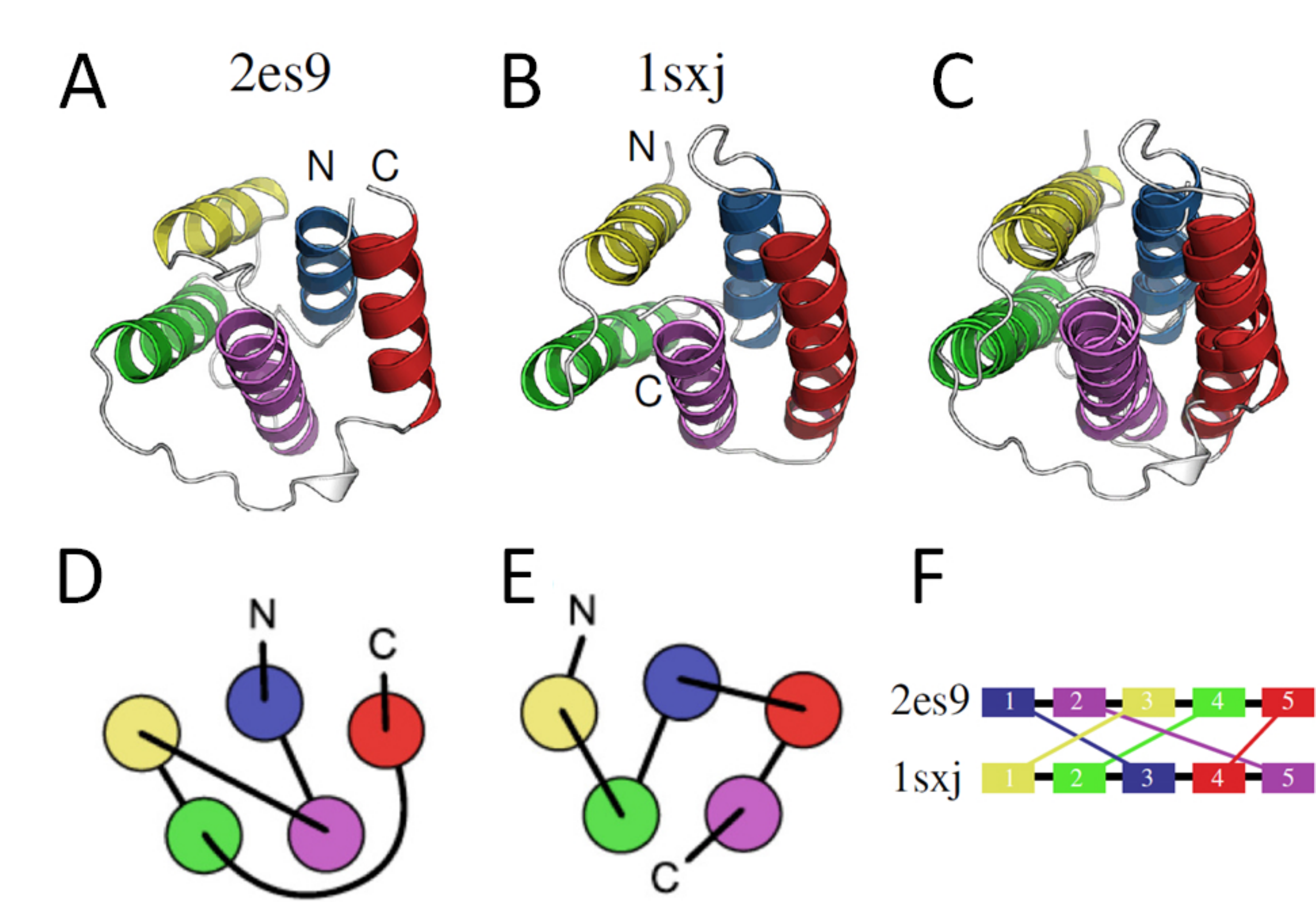}
\caption{
An example of a non-sequential structure alignment. A) Structure of Q8ZRJ2 (PDB code:\,2es9), B) structure of the eukaryotic clamp loader (PDB code:\,1sxj), and C) the superimposition of Q8ZRJ2 and the eukaryotic clamp loader obtained by the non-sequential alignment program MICAN \cite{Minami13}. In A--C, the structurally equivalent regions are drawn with the same color. It can be  clearly seen that all  helices are well superimposed if both the chain direction and the connectivity are ignored.  D and E are two-dimensional diagrams of protein topology of Q8ZRJ2 (A) and eukaryotic clamp loader (B), respectively. F) Correspondence relation of helices obtained by MICAN. Reproduced from \cite{Minami_PhD} with permission.
}
\label{Fig:MICAN}
\end{figure}

At a larger scale, local cooperative structures, foldons or modules, are assembled into the native conformation  in a further cooperative way. A question in this scale is how such  long-range cooperative assembly is realized. Here, the geometrical analysis  sheds light on this problem. One of the present authors developed an efficient non-sequential structure alignment software, MICAN \cite{Minami13}, and demonstrated that the spatial arrangement of SSEs of numerous different proteins can be precisely  superposed on each other if we disregard both the chain direction in SSEs and the manner those SSEs  are connected by chains \cite{Minami13,Minami14}. An example of a non-sequential structure alignment by MICAN is shown in Fig.~\ref{Fig:MICAN}. Indeed,  approximately 80~\% of the fold representatives defined in the SCOP database \cite{Andreeva08} share the same spatial arrangement of SSEs with other folds \cite{Minami14}. Because it is widely accepted that proteins with different folds are very unlikely to be evolutionarily related, this frequent sharing of the same SSE arrangement suggests that  particular SSE arrangements were evolutionarily selected as liquid-crystal-like configurations, which satisfy the  chemical or physical requirements for interactions. With the same SSE arrangement,  the non-local interactions in native conformations can be similarly stable, but local interactions can exhibit significantly different stabilities, depending on the connectivity of the SSEs. In addition, differences in the chain connectivity can modify  the  entropy reduction process along the folding funnel. To elucidate the relative importance of local versus non-local interactions as well as the role of entropy in the SSE assembly, it would be interesting to compare folding pathways for a set of proteins that share the same SSE arrangement but have different topologies. For such a purpose, the WSME model would play an important role, as implicated by the successful description of the folding pathways of both the wild type and the circular permutant of DHFR \cite{Inanami14}.

Conclusively, we address the implications of the coarse-grained modeling studies discussed in this review. Protein folding is a complex molecular process, affected by various atomic interactions; non-native interactions, particularly  non-native disulfide bonds,  slow down the folding process.  Isomerization of proline or other residues affects  the folding/unfolding rates. Cooperative exclusion of water molecules and the concomitant hydrophobic  packing in each local part affect the height and position of the barrier in the free-energy landscape of folding. Some of these features, such as the effects of non-native interactions and proline isomerization, have been explicitly considered in the kinetic description using the WSME model \cite{Inanami14}. 
Here, we emphasize that important aspects of these atomic features are represented in a coarse-grained way, which are compatible with the core assumption of the WSME model that is the cooperativity in forming local structural modules and assembling  those local structures, as indicated by the agreement between the WSME results and the observed data. Therefore, the analyses of  modularity and cooperativity  with the WSME model provide guidelines on how to represent the effects of atomic interactions  in a coarse-grained way to construct models of complex problems, such as allostery dynamics \cite{Terada13}. Therefore, coarse-graining methods should  provide insights on protein evolution, development of techniques for protein structure prediction, and protein engineering. Finally, this approach using simplified statistical mechanical models, which was pioneered by Sait$\hat{\rm o}$, should continue to play an important role in this modern field of protein biophysics.

\section*{Acknowledgment}
This study was supported by JSPS KAKENHI Grant Number JP16H02217, CREST of the Japan Science and Technology Agency, and Riken Pioneering Project ``Cellular Evolution''.

\section*{Conflicts of Interest}
The authors declare no competing financial interest.

\section*{Author Contribution}
M.S., G.C., and T.P.T.  co-wrote the manuscript.

\end{document}